\documentclass[aps,prb,preprint,showpacs,floatfix,superscriptaddress]{revtex4-1}
\usepackage{graphicx,epsfig,dcolumn,amssymb,amsmath}
\usepackage{booktabs}
\usepackage{float}
\usepackage{color}
\usepackage{verbatim}
\begin{document}

\title{CsPbBr$_{3}$ Perovskites: Theoretical and Experimental Investigation on Water-Assisted Transition From Nanowire Formation to Degradation}

\author{B. Akbali}
\affiliation{Department of Physics, Izmir Institute of Technology, 35430, Izmir, Turkey}

\author{G. Topcu}

\affiliation{Department of Materials Science and Engineering, Izmir Institute of Technology, 35430, Izmir, Turkey}

\author{T. Guner}

\affiliation{Department of Materials Science and Engineering, Izmir Institute of Technology, 35430, Izmir, Turkey}

\author{M. Ozcan}

\affiliation{Department of Photonics, Izmir Institute of Technology, 35430, Izmir, Turkey}

\author{M. M. Demir}
\affiliation{Department of Materials Science and Engineering, Izmir Institute of Technology, 35430, Izmir, Turkey}

\author{H. Sahin}
\email{hasansahin@iyte.edu.tr}
\affiliation{Department of Photonics, Izmir Institute of Technology, 35430, Izmir, Turkey}
\affiliation{ICTP-ECAR Eurasian Center for Advanced Research, Izmir Institute of Technology, 35430, Izmir, Turkey}

\date{\today}
\pacs{84.60.Jt, 71.15.Mb, 81.05.ue, 81.20.-n}
\date{\today}

\begin{abstract}

Recent advances in colloidal synthesis methods have led to increased research focus on halide perovskites.  Due to highly ionic crystal structure of perovskite materials, stability issue pops up especially against polar solvents such as water. In this study, we investigate water-driven structural evolution of  CsPbBr$_{3}$  by performing experiments and state-of-the-art first-principles calculations. It is seen that while optical image shows the gradual degradation of yellowish-colored CsPbBr$_{3}$ structure under daylight, UV illumination reveals that the degradation of crystals takes place in two steps; transition from blue-emitting to green-emitting structure and and then transition from green-emitting phase to complete degradation. We found that as-synthesized CsPbBr$_{3}$ NWs emit blue light under 254 nm UV source and before the degradation, first CsPbBr$_{3}$ NWs undergoes a water-driven structural transition to form large bundles. It is also seen that formation of such bundles provide longer-term environmental stability. In addition theoretical calculations revealed how strong is the interaction of water molecules with ligands and surfaces of CsPbBr$_{3}$ and provide atomistic-level explanation to transition from ligand-covered nanowires to bundle formation. Further interaction of green-light-emitting bundles with water causes complete degradation of CsPbBr$_{3}$ and photoluminescence signal is entirely quenched. Moreover, Raman and XRD measurements revealed that completely degraded regions are decomposed to PbBr$_{2}$ and CsBr precursors. We believe that findings of this study may provide further insight into the degradation mechanism of CsPbBr$_{3}$ perovskite by water.

\end{abstract}

\maketitle

\section{Introduction}

Halide perovskites, having the structure of ABX$_{3}$ (where A is organic: CH$_{3}$NH$_3^+$ (MA), HC(NH$_{2}$)$_2^+$ (FA) or inorganic: Cs$^+$ cation, B is metal cation: Pb$^{2+}$, Sb$^{2+}$, Sn$^{2+}$, and X is halide anion: Cl$^-$, I$^-$, Br$^-$), have been known since 1950s.\cite{moller1958} With the discovery of MAPbI$_3$ as photosensitizer in dye-sensitized solar cells (DSSCs) at 2009s,\cite{kojima2009} these materials have started to receive more attention.\cite{stoumpos2015} Since then, in a short period of time, perovskite solar cells have become able to improve the conversion efficiency from 3.81\% to almost 20\%.\cite{kojima2009,liu2014perovskite,zhou2014,jeon2015} Apart from the success of these materials in photovoltaic applications\cite{im20116}, it has been demonstrated recently that these materials can be applied also to light-emitting diodes (LEDs),\cite{cho2015,byun2016,li201750} lasers,\citep{zhu2015,yakunin2015low} photodetectors,\cite{ramasamy2016all} etc. due to their unique optical properties; high quantum yield ($90\%$), wavelength tunability, and color purity.\cite{li2017,bai2016,stoumpos2016,li2016,protesescu2015,kovalenko2017}

Even though all-inorganic perovskites are better in terms of intrinsic stability than the organometallic halide ones, stability is still a challenge especially against moisture and polar solvents such as water, ethanol, acetone, etc.\cite{li2017,huang2017,iso2018,kovalenko2017} Such high chemical instability mainly stems from high ionic character of the compounds. However, practical applications require deeper understanding of degradation mechanisms for synthesis of highly stable halide perovskites under ambient conditions. To date, various approaches have been developed such as mixing with mesoporous silica,\cite{wang2016} core-shell structure,\cite{bhaumik2016,li2017cs,qiao2017} different surface treatments (other than usual ones; Oleic acid (OA), and Oleylamine (OAm))\cite{huang2016,luo2016} or encapsulation with polymers\cite{wang2016encap,raja2016,wei2017} to obtain perovskite nanocrystals with high stability. 

Recently, among halide perovskites, all-inorganic CsPbX$_{3}$ nanocrystals have started to draw much attention because of their high photoluminescence quantum yields and controllable morphologies.\cite{sun2016ligand,zhang2016enhancing,swarnkar2015colloidal,xue2017constructing} Chen \textit{et al.} reported efficient synthesis technique to prepare CsPbX$_{3}$ nanocrystals with tunable composition, luminescence characteristics, and morphologies.\cite{chen2017solvothermal}

Degradation mechanisms at atomic-level and possible stabilization techniques are still open questions in the growing field of perovskite nanocrystals. A very recent study of Yuan \textit{et al.} reported that both light and humidity may degrade CsPbI$_{3}$ quantum dots.\cite{yuandegradation} Recent perspective study on lead halide perovskite solar cells (PSCs) enlightens the defect tolerance and stability of the material.\cite{huang2017lead} Lejitas and co-workers reviewed strategies to overcome  the issues of structural, thermal, and atmospheric degradation of CsPbX$_{3}$ nanocrystals.\cite{leijtens2017towards} In addition to experimental studies, tremendous efforts have been also performed on PSCs by carrying out density functional theory (DFT) calculations.\cite{haruyama2015first,kawai2015mechanism,geng2014first,yin2015halide,iyikanat2017thinning} Kang \textit{et al.} theoretically predicted that regarding its electronic properties CsPbBr$_{3}$ is a defect-tolerant semiconductor.\cite{kang2017high} In addition, theoretical studies on defects in perovskites have been widely studied.\cite{yin2014unusual,kim2014role,azpiroz2015defect,haruyama2017first} Although the organic-inorganic hybrid lead halide perovskites have been studied theoretically and experimentally, the pure inorganic alternative, CsPbBr$_{3}$ phase, has been recently found to possess most of the good properties of the hybrid lead halide counterpart.\cite{protesescu2015,bekenstein2015highly}

The paper is organized as follows: Detailed information about computational and experimental methodologies are given in Section II. Characteristic properties of as-synthesized CsPbBr$_{3}$ are investigated in Sec. III. Water-assisted transition from blue to green emitting structure is explained in Sec. IV and complete degradation of the perovskite material is discussed in Sec. V. Finally, we conclude our results in Sec. VI.

\begin{figure*}
\includegraphics[width=16 cm]{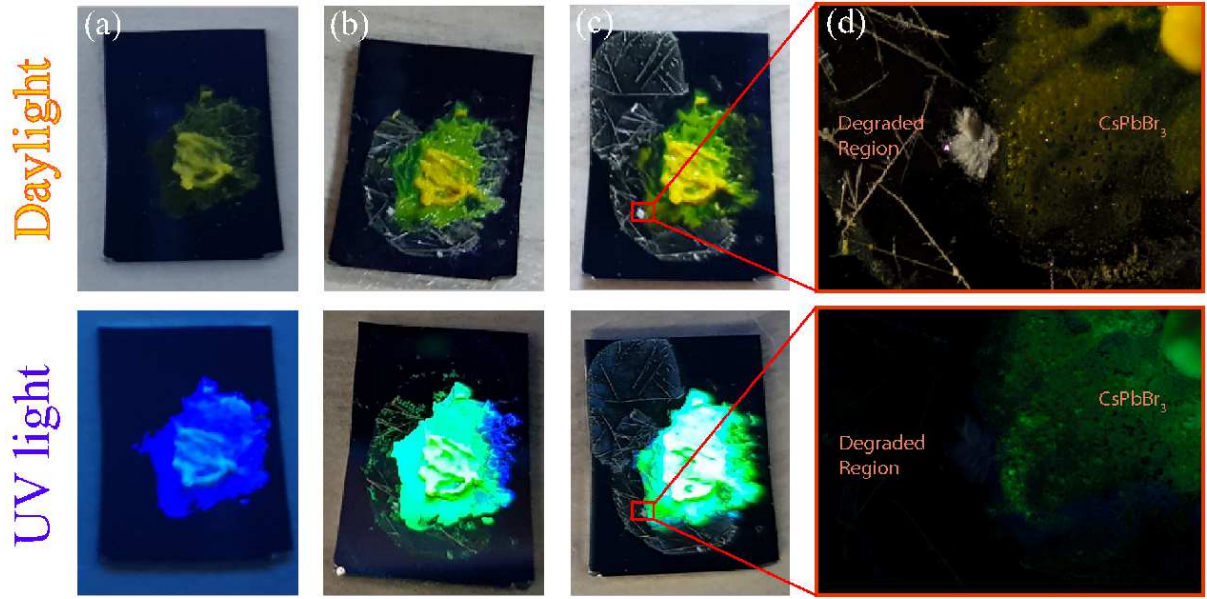}
\caption{\label{structure1}
(Color online) 
Appearance of CsPbBr$_3$ under daylight and UV illumination; (a) initially after casting, (b) after 24 and (c) 144 hours treated with water, respectively. Focused region in (c) presents the optical microscope image of related sample under daylight and UV excitation.}
\end{figure*}

\section{Methodology}

\subsection{Experimental Methodology}

%\subsubsection{Materials}
Cesium carbonate (Cs$_2$CO$_3$, 99.9\%, Sigma-Aldrich), lead(II) bromide (PbBr$_2$, $\geq$98\%, Sigma-Aldrich), oleic acid (OA, 90\%, Alfa Aesar), oleylamine (OLA, 90\%, Sigma-Aldrich), 1-octadecene (ODE, 90\%, Sigma-Aldrich), dimethylformamide (DMF, $\geq$99.9\%, Tekkim), Hexane ($\geq$98\%, Sigma-Aldrich) and acetone (Merck , $\geq$99.5\%) were purchased and used as received without any further purification. Oxidized silica substrate was purchased from University Wafers.

%\subsubsection{Synthesis}
\textit{Synthesis of Cs-oleate:} Cs-oleate solution was synthesized with slight modifications by following the Amgar et al.\cite{amgar2017} Cs$_2$CO$_3$ (0.2 g), OA (625$\mu$L) and ODE (7.5 mL) were loaded to 3 necked flask, and dried under vacuum (150 mbar) at $120^{\circ}$C for 1h. Subsequently, mixture was heated to $150^{\circ}$C under N$_{2}$, and reaction was maintained until all Cs$_{2}$CO$_3$ consumed by OA. Afterwards, yellowish Cs-oleate solution was gradually cooled down (it has to be pre-heated to $100^{\circ}$C before using).

\textit{Synthesis of CsPbBr$_3$:} CsPbBr$_3$ crystals were prepared in four steps with slight modifications based on the procedure.\cite{amgar2017} An aliquot of 0.125 mL OA, 0.125 mL OLA and 1.25 ml ODE were loaded to glass vial. Subsequently, 0.1 mL of pre-heated Cs-oleate solution was added to mixture, and addition of 0.2 mL of PbBr$_2$ precursor solution (0.4 M, heated for 1 h at $80^{\circ}$C until full dissolution) followed it. After 10 s, 5 mL of acetone were rapidly added to trigger the crystallization of the CsPbBr$_3$. Stirring was maintained for 30 minutes and green precipitates were collected by using centrifuge (6000 rpm, 10 m). Precipitates were re-dispersed in Hexane.

\textit{Preparation of Water Contact CsPbBr$_3$ Surface:} First, 150 $\mu$L CsPbBr$_{3}$/Hexane dispersion was cast on an oxidized silica substrate (approximately 1 cm$^2$). CsPbBr$_{3}$ structures were formed immediately after the quick evaporation of hexane. Second, an aliquot of distilled water was put over the CsPbBr$_{3}$ coated silica substrate and waited till the water completely evaporates. Characterizations were carried out in ambient conditions. For further aging of the crystals, water driven transitions were conducted by adding desired amount of water. Water contact time was recorded as the total time of exposure.

%\subsubsection{Characterization} 
The diffraction profile of the CsPbBr$_3$ structures was recorded with an X-ray diffractometer (XRD, X’Pert Pro, Philips, Eindhoven, the Netherlands). Scanning electron microscopy (SEM; Quanta 250, FEI, Hillsboro, OR, USA) was used to determine CsPbBr$_3$ morphology in back-scattering electron (BSE) detectors. Image of the degraded crystals were captured via optical microscope (BX 53, Olympus, Tokyo, Japan). Emission spectra was determined by USB2000+ spectrometer (Ocean Optics Inc., Dunedin, FL, USA) via a premium fiber cable. Raman (Horiba Xplora plus) was used to determine fingerprint Raman-active vibrations of CsPbBr$_{3}$ structures. Absorption was collected via using OLYMPUS (CX-31) optical microscope integrated with USB2000+ spectrometer.

\subsection{Computational Methodology}

 To investigate interaction between Cs- and Pb-rich surfaces of orthorhombic CsPbBr$_{3}$ with water, OA, and OAm molecules, we performed density functional theory-based calculations using the projector augmented wave (PAW)\cite{kresse1999ultrasoft,blochl1994projector} potentials as implemented in the Vienna \textit{ab initio} Simulation Package (VASP).\cite{kresse1993ab,kresse1996efficient} The local density approximation (LDA)\cite{perdew1981self} was used with the inclusion of spin-orbit coupling (SOC) to describe the exchange and correlation potential as parametrized by the Ceperley and Alder functional to describe the exchange and correlation potential.\cite{ceperley1980ground} Bader technique was used to analyze the partial charge transfer on the atoms.\cite{henkelman2006fast} 
 
 A plane-wave basis set with kinetic energy cutoff of 500 eV was used for all the calculations. The total energy difference between the sequential steps in the iterations was taken to be 10$^{-5}$ eV for the convergence criterion. The total force in the unitcell was reduced to a value of less than 10$^{-4}$ eV$/$\AA{}. $\Gamma$-centered k-point meshes of 3 $\times$ 3 $\times$ 3 were used. A vacuum space of 10 \AA{} was incorporated to avoid interaction with adjacent surfaces. Gaussian smearing of 0.1 eV was used for electronic density of states calculations. Spin-polarized calculations were performed in all cases.

\section{Results}

Structural and electronic evolution of the water interacting CsPbBr$_3$ crystals were captured under both daylight and UV light (254 nm) at different times, and presented in Fig. \ref{structure1}. First, silica substrates were identified as neat showing black color under both daylight and UV. After dropping the CsPbBr$_3$ / hexane dispersion over the substrate, the sample was observed to become green and yellowish like color under daylight and explicit blue under UV. After interacting with water molecules about 24 hours, as shown in Fig. \ref{structure1} (b), CsPbBr$_3$ turn into explicit greenish color with some white crystals move around expanding the sample volume, which exhibits green emission covering large area while leaving a small region as emitting blue under UV light. Further interaction with water leads to formation of white and relatively large crystals that do not exhibit luminescence under UV illumination due to possible degradation (Fig. \ref{structure1} (c) and (d)). Monitoring the degradation under UV light reveals that degradation occurs in two different steps: formation of green-emitting phase and complete degradation. Therefore, following chapters are devoted to experimental and atomic-level understanding of degradation of CsPbBr$_{3}$ crystal by water.

\subsection{Characteristic Properties of Blue Light Emitting C\lowercase{s}P\lowercase{b}B\lowercase{r}$_{3}$}

\begin{figure}
\includegraphics[width=14 cm]{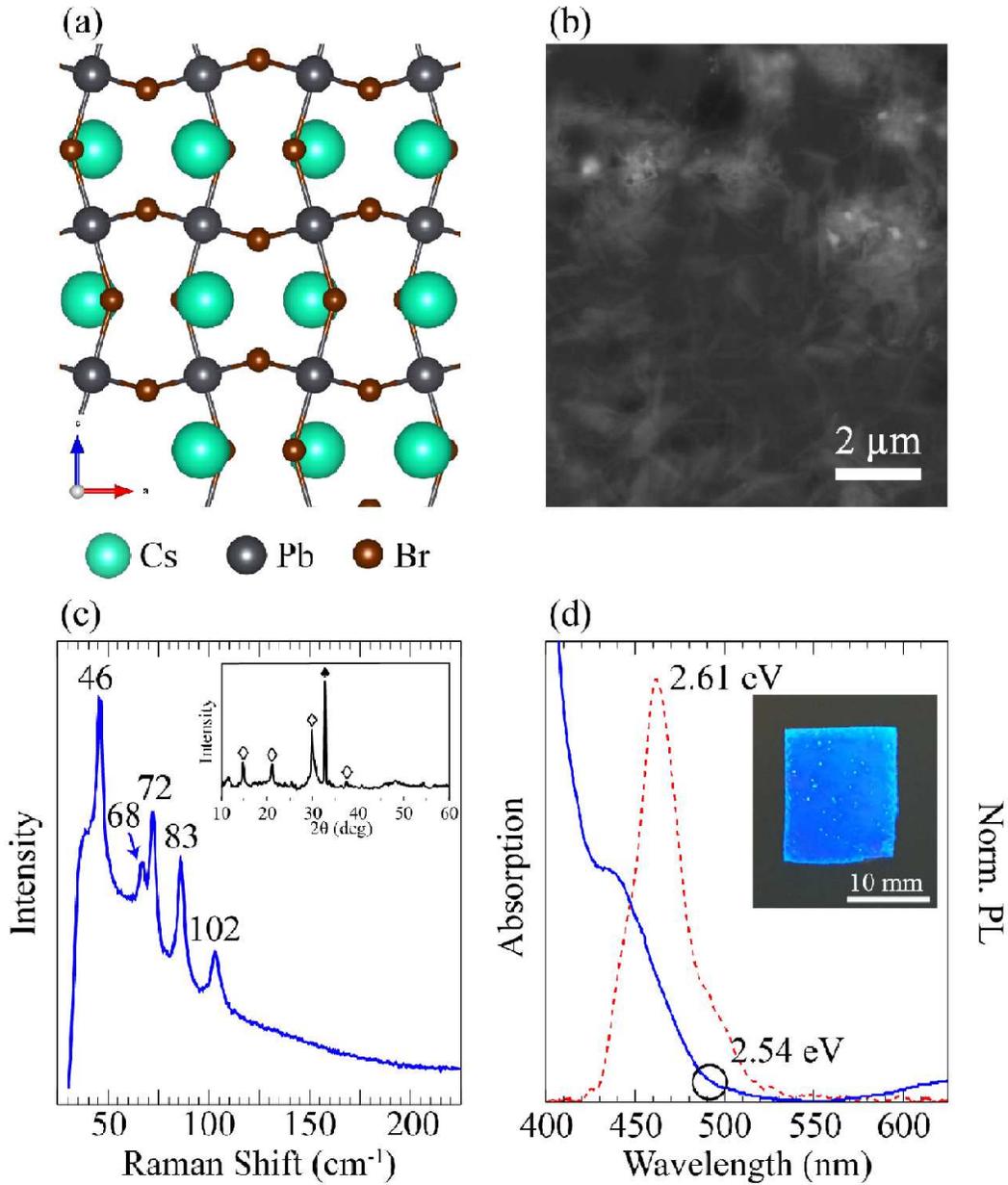}
\caption{\label{structure2}
(Color online) 
(a)Crytal structure, (b) SEM image, (c) Raman measurement (inset; XRD pattern) and (d) photoluminescence and absorption spectra (inset; photograph under 254 nm UV light) of CsPbBr$_3$ NWs.}
\end{figure}

CsPbX$_{3}$ nanocrystals exhibit three different structural phases: cubic (Pm-3m), orthorhombic (Pnma), and tetragonal (P4/mbm).\cite{smith2015interplay,wang2015pressure,cottingham2016} At room temperature, CsPbBr$_{3}$ has been shown to possess a thermodynamically preferred orthorhombic structure.\cite{zhang2015solution} In our calculations, structural properties of orthorhombic CsPbBr$_{3}$ is investigated, as seen in Fig. \ref{structure2} (a). Structural analysis reveals that optimized lattice parameters of bulk CsPbBr$_{3}$ is \textit{a} = 8.34 \AA{}, \textit{b} = 7.89 \AA{}, and \textit{c} = 11.29 \AA{}. Each Br atom bonds with two Pb atom with a bond length of 2.92 \AA{}. Br-Pb-Br bond angle varies between 85$^{\circ}$ and 90$^{\circ}$. Bader charge analysis shows that each Br atom receives 0.6\textit{e}/atom from Cs (0.8\textit{e}/atom) and Pb (1\textit{e}/atom) atoms. Besides, it is seen that the bond between Pb and Br atoms has strong ionic character. On the other hand, Cs atoms slightly bind to the other atoms in the system even though the system receives charge from Cs atoms. 

SEM image presented in Fig. \ref{structure2} (b) shows the morphological characteristics of CsPbBr$_3$ crystals. It is observed that the crystals, even though they have negligible aggregation, possess 1-dimensional shape. By selecting the individual ones from the image reveals that these NWs have nano scale diameter ($\sim$ ~50 nm) and submicron lengths ($\sim$ 0.5-1.5 $\mu$m).

Raman measurement performed at room temperature with 785 nm laser excitation shows that CsPbBr$_3$ has five Raman-active modes as presented in Fig. \ref{structure2} (c). The vibration of the metal-halide sublattice, prominent peak of CsPbBr$_{3}$, is measured at 72 cm$^{-1}$. According to a previous Raman study of the CsPbCl$_{3}$ crystal with \textit{Pnma} phase,\cite{calistru1997identification} peak at 72 cm$^{-1}$ is assigned to the vibrational mode of [PbBr$_{6}$]$^{4-}$ octahedron and motion of Cs$^{+}$ cations. Moreover, it is seen that additional peaks appear at 46, 68, 83 and 102 cm$^{-1}$. 

For further information on the crystal structure, XRD pattern of CsPbBr$_3$ NWs is also presented in the inset of Fig. \ref{structure2} (c). Diffraction pattern is in good agreement with the standard orthorhombic phase as the crystal structure of CsPbBr$_3$, where \textit{2$\theta$} at $15^{\circ}$, $22^{\circ}$, $30^{\circ}$, and $31^{\circ}$ reflections, marked with tile symbol, correspond to the (110), (020), (004), and (220), respectively.\cite{cottingham2016} The peak at $33^{\circ}$ is due to silicon wafer. Among the reflections, asymmetry between (004) and (220) planes indicates a good morphological support in the sense of producing NW geometry. 

Optical bandgap of CsPbBr$_{3}$ NWs was determined via absorption spectrum, as shown in Fig. \ref{structure2} (d). Based on the data presented in Fig. \ref{structure2} (d), NWs show broad range of absorption starting from the wavelength of $\sim$488 nm, which follows an increasing trend with the decreasing wavelength. Absorption rate grows almost exponentially below $\sim$425 nm. To estimate bandgap, wavelength of where the absorption begins is considered, which gives rough value about 2.54 eV. The bandgap, which corresponds to a wavelength of 488 nm is verified by photoluminescence (PL) spectrum, which is given with a dashed line in Fig. \ref{structure2} (d), of the NWs. In that spectrum, it is observed that NWs have narrow (FWHM = 38 nm) blue emission with maximum PL intensity at 475 nm under 254 nm UV light. The wavelength, which corresponds to 2.61 eV, is very close to the estimated bandgap value above. For visualization, photograph of the casted CsPbBr$_3$ NWs over silica substrate under UV illumination, which is an explicit blue, is presented as the inset of Fig. \ref{structure2} (d).

\subsection{Water-assisted Transition from Blue to Green Light Emitting Structure}

\begin{figure}
\includegraphics[width=14 cm]{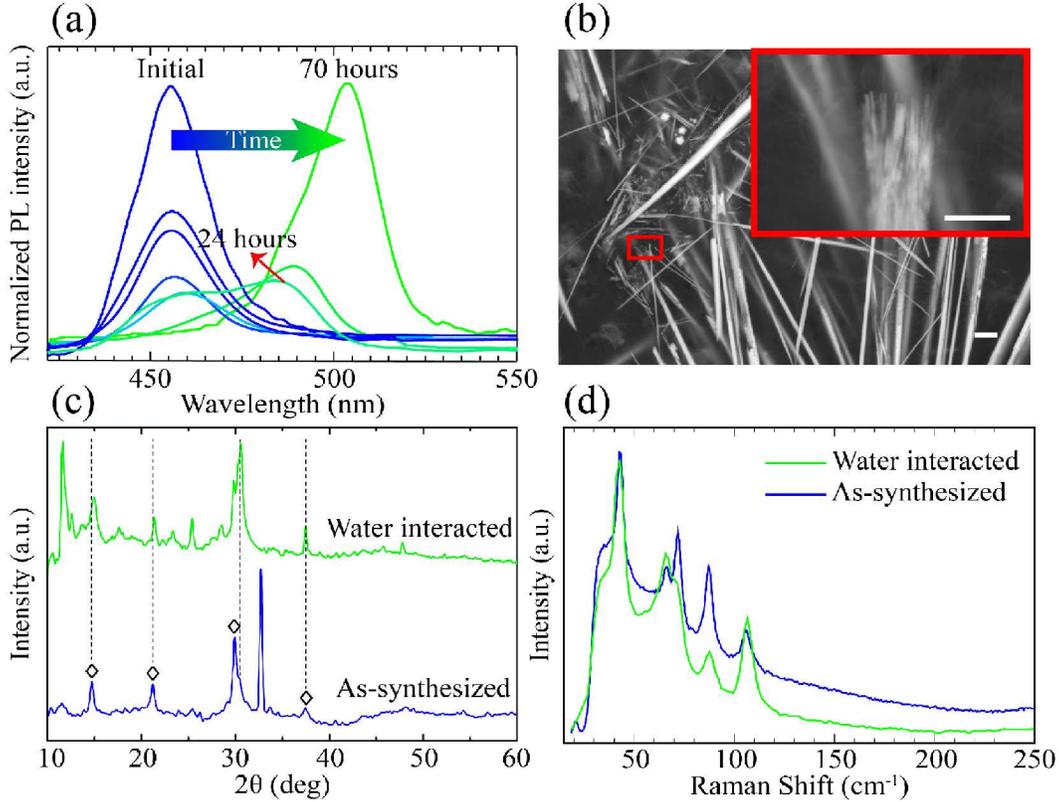}
\caption{\label{structure3}
(Color online) 
(a) Time-dependent photoluminescence of as-synthesized NWs interacting with water, (b) SEM image of the resulting NW structure after 24 hours of interaction , and (c)-(d) are xrd measurement and Raman spectra of both as-synthesized and water interacted samples. Scale bars are 2 $\mu$m.}
\end{figure}

While CsPbBr$_3$ preserves its yellowish color during the water treatment, water-driven transition into the green-emitting phase can be observed under UV. As shown in Fig. \ref{structure3} (a), the transition from blue to green emission due to water was recorded with different times. After 24 hours, emission becomes consisting of two distinct signals; one signal around 450 nm represents the individual NWs (verifies the blue region in Fig. \ref{structure1} (b)) and a signal around 500 nm (greenish), which indicates a significant red-shift. From 24 hours to 70 hours, blue signal reduces and finally disappears while greenish signal increases and dominates as a single signal at last.

SEM image shown in Fig. \ref{structure3} (b) reveals formation of larger crystals that were grown particularly in longitudinal direction compared to individual NWs, reaching $>$5 $\mu$m. Inset demonstrates an image that was taken with higher magnification over the edge of one of these crystals. It was observed that the tip of a large and rod-like crystal consists of many NWs, and inset of Fig. \ref{structure3} (b) verifies the bundle formation through individual NWs. Therefore, the water-driven red-shift in emission clearly stems from quantum size effect, which is led by the structural transition from nanowire to bundle. 

Crystal structure of green-emitting CsPbBr$_3$ was determined by x-ray diffraction measurement, as shown in Fig. \ref{structure3} (c). It is seen from the \textit{2$\theta$} reflections at $15^{\circ}$, $22^{\circ}$, $30^{\circ}$, and $31^{\circ}$ confirms that blue and green-emitting phases correspond to the same crystal structure. However, emergence of additional reflections \textit{2$\theta$} at $12^{\circ}$ and $25^{\circ}$ are indication of locally formed Cs$_4$PbBr$_6$.\citep{zhang2017zero} Raman spectra of green emitting CsPbBr$_3$ bundles shows that prominent peak at 72 cm$^{-1}$ and the other 4 modes still exist, as presented in Fig. \ref{structure3} (d). Raman activity of water interacted region is the same with as-synthesized CsPbBr$_3$ crystals. Therefore, vibrational characteristics of CsPbBr$_3$ NWs remains unchanged through the bundle formation.

\begin{figure}
\includegraphics[width=8.5 cm]{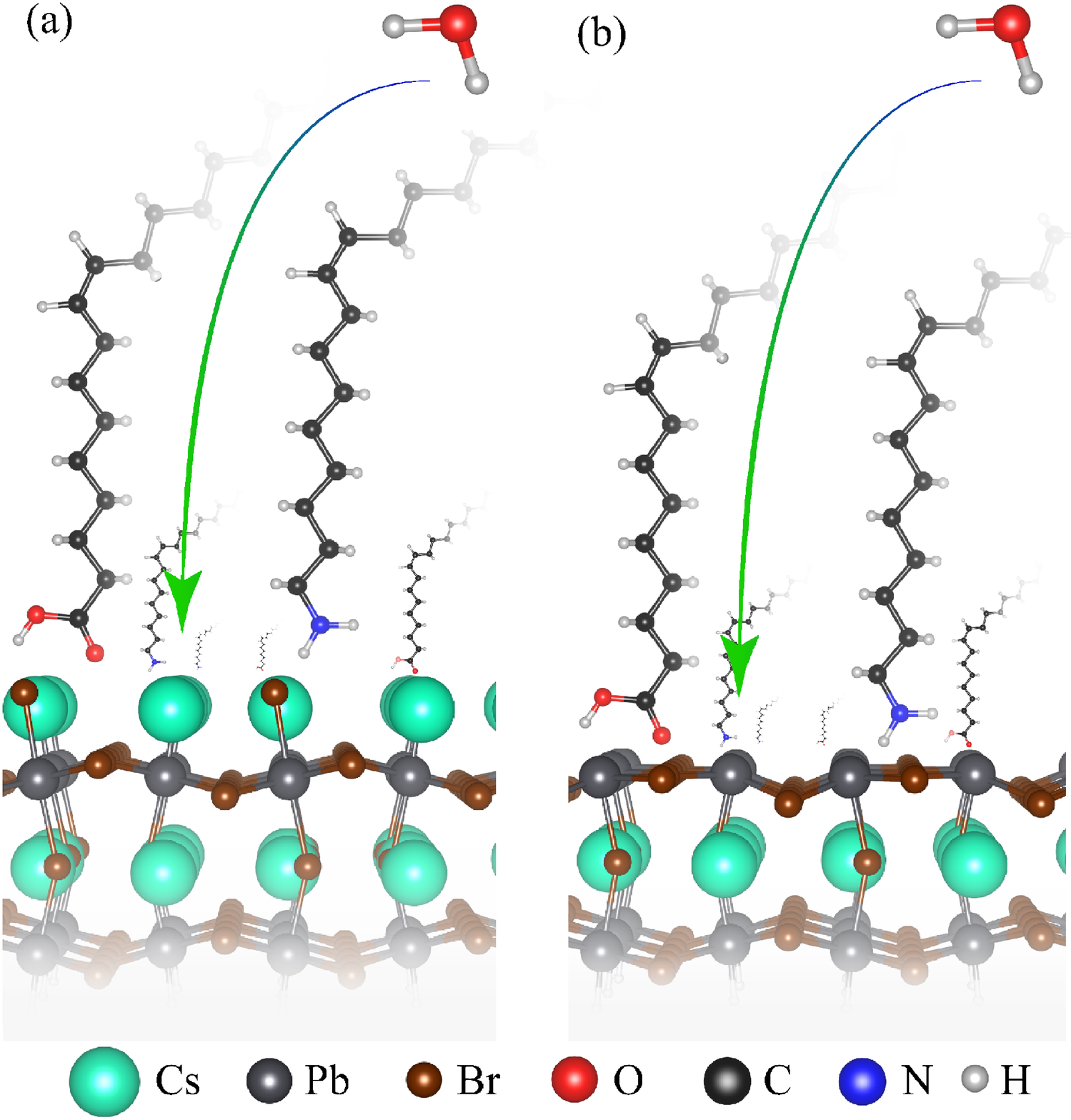}
\caption{\label{structure4}
(Color online) 
Side views (a) Cs-terminated and (b) Pb-terminated CsPbBr$_{3}$ surface.}
\end{figure}

Here, for deeper understanding of the water-driven nanowire-to-bundle transformation, we employ state-of-the-art first principles calculations. In order to examine all possible surfaces, we truncated bulk orthorhombic CsPbBr$_{3}$ at Cs and Pb surfaces as shown in Fig. \ref{structure4} (a) and (b). Both surfaces are saturated with H atom to evade possible magnetization in the system. 

For the surface cut from Cs atoms, bond angle between Pb and Br atoms is calculated to be the same as bulk form. Due to the surface relaxation Pb-Br bond length varies between 2.86-2.94 \AA{}. In addition, Bader charge analysis reveals that Pb atom donate (1.1\textit{e}) to Br atoms (0.5-0.6\textit{e}). Pb atoms donate 0.1\textit{e} more comparing whit bulk CsPbBr$_{3}$. For Pb truncated surface, it is clearly seen that there is a surface reconstruction which Pb-Br atoms make a line ($\sim$ $180^{\circ}$), as shown in Fig. \ref{structure4} (b). Bond length Pb-Br atoms vary between 2.82 and 2.95 \AA{} which is wider range than Cs-surface. Bader charge analysis of Pb-terminated surface reveals that while Pb atoms donate 1.1\textit{e}, Br atoms receive 0.5-0.6\textit{e}.

Total energy optimization calculations for relaxation of H$_{2}$O molecules on two different surfaces show that water molecules prefer to bind on the bridge site on Pb-Br bond. In addition, H$_{2}$O molecules bind to both surfaces via lone pairs of O atoms. For Cs-terminated surface, bond length between O and Cs atoms is calculated to be 2.91 \AA{}. Charge analysis shows that H$_{2}$O molecule receives 0.2\textit{e} from Cs atom. Binding energy (E$_{b}$) of the molecule is calculated to be 649 meV. H$_{2}$O molecule bind to Pb atom with a bond length of 2.53 \AA{}. The molecule prefers to bind the surface through O atom. without charge transfer between the molecule and the surface. On this surface, the binding energy (E$_{b}$) of the molecule is calculated to be 808 meV. Apparently, H$_{2}$O molecules strongly interact with the different surfaces of CsPbBr$_{3}$.

We also calculate the interactions between surface ligands, OA and OAm, with the NW surfaces. Optimized structures of the ligands are shown in Fig. \ref{structure4}.  On the Cs-terminated surface, E$_{b}$ of OA is calculated to be 698 meV. On the contrary, E$_{b}$ of water molecule on the same surface is calculated as 649 meV and it is comparable with OA' binding energy. In addition E$_{b}$ of OAm on the same surface is found to be 463 meV.  For Pb-terminated surface, binding energy of OAm is calculated to be 874 meV. It is comparable with H$_{2}$O molecule's binding energy (808 meV) on the same surface. Furthermore, binding energy of OA molecule is found to be 221 meV. 

To provide a complete discussion, interaction of water with ligands is also taken into account. It is found that E$_{b}$ of H$_{2}$O with OA and OAm is calculated to be 898 and 550 meV, respectively. It can be concluded that oleic acid is more likely to bind with water molecule than any other surface of CsPbBr$_{3}$. So H$_{2}$O molecules mediate the detachment of OA molecules from the surface of the NWs. Hence, minor phase transformation from CsPbBr$_{3}$ to Cs$_{4}$PbBr$_{6}$ may arise from lack of OA that leads to excessive OAm on the surface of the NWs.\cite{thumu2017mechanistic} 

As a result, the binding energies of water and ligands on the surface of CsPbBr$_3$, which are comparable to each other, reveal that water is responsible from the removal of ligands over the CsPbBr$_3$ surface. It appears that detached ligands on the surface of NWs yield to formation of bundles composed of individual NWs.

\subsection{Complete Degradation}

\begin{figure}
\includegraphics[width=14 cm]{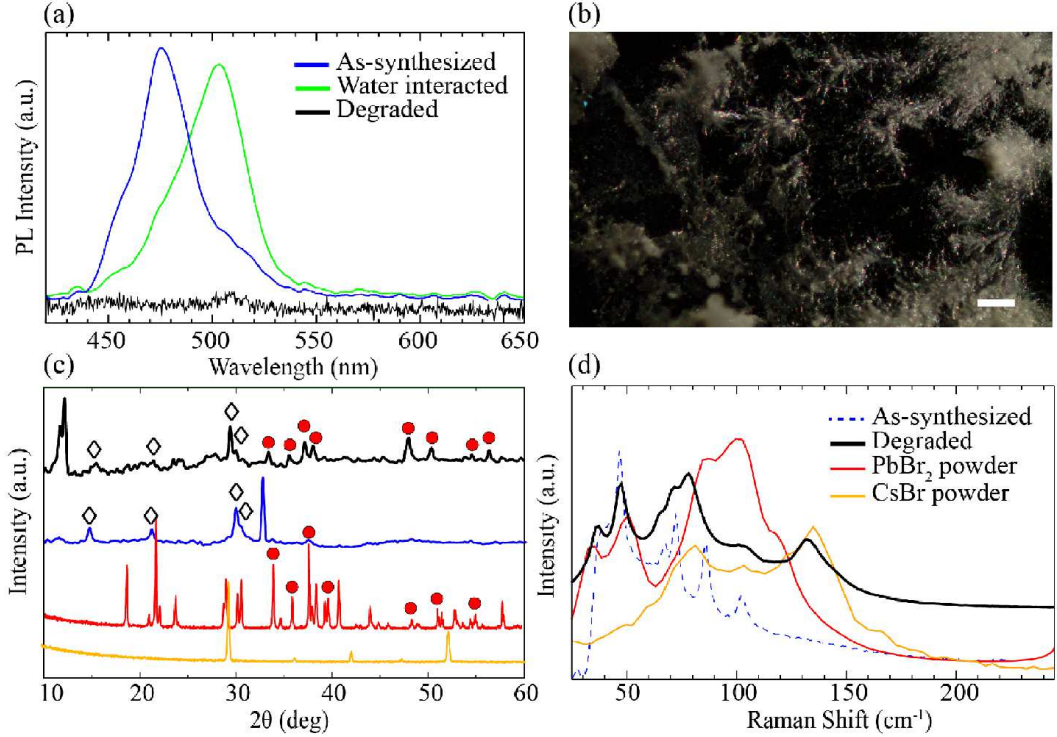}
\caption{\label{structure5}
(Color online) 
(a) PL measurement of CsPbBr$_{3}$ NW to degraded crystals, (b) Optical image from non-emitting bundles, and (c) X-ray diffraction pattern and (d) Raman measurement of CsPbBr$_{3}$ NWs, degraded crystals, CsPb and PbBr$_{2}$ powder. Scale bar is 50 $\mu$m.}
\end{figure}

This section is devoted to understanding of how the green light emitting CsPbBr$_{3}$ bundles interact with water and become completely degraded. As shown in Fig. \ref{structure5} (a), further water treatment of green-emitting phase having intense PL signal at 500 nm results in transformation into another phase that has no optical activity. Optical image presented in Fig. \ref{structure5} (b) reveals that formation of these non-emitting crystal structures is accompanied by formation white colored domains.

For understanding of the final structure in terms of crystallographic perspective, XRD measurements were employed to white crystals, namely non-emitting large bundles. XRD patterns of degraded and as-synthesized NWs, CsBr, and PbBr$_{2}$ are presented in Fig. \ref{structure5} (c). Identical reflections ($15^{\circ}$ and $30^{\circ}$) of perovskite, which indicate (110) and (004) planes, is still observed. The reflections of degraded form may be assigned to residual NWs. However, according to additional reflections of degraded crystals, it can be said that crystallographic nature shows alteration as much as transformation shows in morphology. Nevertheless, the additional reflections cannot be unambiguously attributed to raw materials (CsBr and PbBr$_{2}$), although several reflections (e.g. $38^{\circ}$, $48^{\circ}$, and $52^{\circ}$) are already matched.

For further analysis of non-emitting regions, we also present the Raman spectrum of degraded regions of the crystal structure in Fig. \ref{structure5} (d). It is seen that the prominent Raman mode of CsPbBr$_{3}$ at 72 cm$^{-1}$ is significantly decreased and is almost disappeared.  Moreover, some novel modes, apparently stemming from the precursors PbBr$_{2}$ and CsBr emerge at 38 and 130 cm$^{-1}$. \cite{isupova1968raman,willemsen1971raman} Therefore, as confirmed by the vibrational spectrum, complete degradation of CsPbBr$_{3}$ from green-emitting phase occurs by turning the material into its constituents.

\section{Conclusions}

In conclusion, we investigated how CsPbBr$_{3}$ perovskite is degraded by water using Raman, XRD,PL measurements and state-of-the-art computational techniques. It is seen that during degradation even though no significant structural changes were visible by the naked eye, UV illumination reveals that the complete degradation takes place in two different steps (i) transformation from nanowires to bundles and (ii) complete degradation from bundles to constituents. As verified by the first-principles calculations, competing interactions between water molecule oleic acid and oleilamine on the surface determine whether the structure crystallizes into a nanowire, bundle or degraded form. Our photoluminescence, Raman and XRD measurements also revealed that during the transition from blue- to green-emitting phase of the CsPbBr$_{3}$ crystal symmetry remains the same. In the final step, complete degradation of CsPbBr$_{3}$ structure takes place by formation of CsPb and PbBr$_{2}$ powders. We believe that these results provide the important advances in understanding the water-driven degradation of perovskite crystals and may construct a theoretical basis for fundamental investigations on their stability.

\section{Acknowledgments}
 Computational resources were provided by TUBITAK ULAKBIM, High Performance and Grid Computing 
 Center (TR-Grid e-Infrastructure). HS acknowledges financial support from the TUBITAK under the project number 117F095.


\begin{thebibliography}{99}

\bibitem{moller1958} C. K. M\O{}LLER, Nature \textbf{182}, 1436 (1958).
 
\bibitem{kojima2009} A. Kojima, K. Teshima, Y. Shirai, and T. Miyasaka, J. Am. Chem. Soc.  \textbf{131}, 6050 (2009).

\bibitem{stoumpos2015} C. C. Stoumpos and M. G. Kanatzidis, Acc. Chem. Res. \textbf{48}, 2791 (2015).

\bibitem{liu2014perovskite} D. Liu and T. L. Kelly, Nature photonics \textbf{8}, 133 (2014).

\bibitem{zhou2014} H. Zhou, Q. Chen, G. Li, S. Luo, T.-b. Song, H.-S. Duan, Z. Hong, J. You, Y. Liu, and Y. Yang, Science \textbf{345}, 542 (2014).

\bibitem{jeon2015} N. J. Jeon, J. H. Noh, W. S. Yang, Y. C. Kim, S. Ryu, J. Seo, and S. I. Seok, Nature \textbf{517}, 476 (2015).

\bibitem{im20116} J.-H. Im, C.-R. Lee, J.-W. Lee, S.-W. Park, and N.-G. Park, Nanoscale \textbf{3}, 4088 (2011).

\bibitem{cho2015} H. Cho, S.-H. Jeong, M.-H. Park, Y.-H. Kim, C. Wolf, C.-L. Lee, J. H. Heo, A. Sadhanala, N. Myoung, S. Yoo,  S. H. Im, R. H. Friend, T.-W. Lee, Science \textbf{350}, 1222 (2015).

\bibitem{byun2016} J. Byun, H. Cho, C. Wolf, M. Jang, A. Sadhanala, R. H. Friend, H. Yang, and T.-W. Lee, Adv. Mater. \textbf{28}, 7515 (2016).

\bibitem{li201750} J. Li, L. Xu, T. Wang, J. Song, J. Chen, J. Xue, Y. Dong, B. Cai, Q. Shan, B. Han, H. Zeng, Adv. Mater. \textbf{29} (2017).

\bibitem{zhu2015} H. Zhu, Y. Fu, F. Meng, X. Wu, Z. Gong, Q. Ding, M. V. Gustafsson, M. T. Trinh, S. Jin, and X. Zhu, Nat. Mater. \textbf{14}, 636 (2015).

\bibitem{yakunin2015low} S. Yakunin, L. Protesescu, F. Krieg, M. I. Bodnarchuk, G. Nedelcu, M. Humer, G. De Luca, M. Fiebig, W. Heiss, and M. V. Kovalenko, Nat. Commun. \textbf{6}, 8056 (2015).

\bibitem{ramasamy2016all} P. Ramasamy, D.-H. Lim, B. Kim, S.-H. Lee, M.-S. Lee, and J.-S. Lee, Chem. Commun. \textbf{52}, 2067 (2016).

\bibitem{li2017} X. Li, F. Cao, D. Yu, J. Chen, Z. Sun, Y. Shen, Y. Zhu, L. Wang, Y. Wei, Y. Wu, H. Zeng, Small \textbf{13}, 1603996 (2017).

\bibitem{bai2016} S. Bai, Z. Yuan, and F. Gao, J. of Mater. Chem. C \textbf{4}, 3898 (2016).

\bibitem{stoumpos2016} C. C. Stoumpos and M. G. Kanatzidis, Adv. Mater. \textbf{28}, 5778 (2016).

\bibitem{li2016} X. Li, Y. Wu, S. Zhang, B. Cai, Y. Gu, J. Song, and H. Zeng, Adv. Funct. Mater. \textbf{26}, 2435 (2016).

\bibitem{protesescu2015} L. Protesescu, S. Yakunin, M. I. Bodnarchuk, F. Krieg, R. Caputo, C. H. Hendon, R. X. Yang, A. Walsh, and M. V. Kovalenko, Nano Lett. \textbf{15}, 3692 (2015).

\bibitem{kovalenko2017} M. V. Kovalenko, L. Protesescu, and M. I. Bodnarchuk, Science \textbf{358}, 745 (2017).

\bibitem{huang2017} H. Huang, M. I. Bodnarchuk, S. V. Kershaw, M. V. Kovalenko, and A. L. Rogach, ACS Energy Lett. \textbf{2}, 2071 (2017).

\bibitem{iso2018} Y. Iso and T. Isobe, ECS J. Solid State Sci. Technol. \textbf{7}, R3040 (2018).

\bibitem{wang2016} H.-C. Wang, S.-Y. Lin, A.-C. Tang, B. P. Singh, H.-C. Tong, C.-Y. Chen, Y.-C. Lee, T.-L. Tsai, and R.-S. Liu, Angew. Chem. Int. Ed. \textbf{55}, 7924 (2016).

\bibitem{bhaumik2016} S. Bhaumik, S. A. Veldhuis, Y. F. Ng, M. Li, S. K. Muduli, T. C. Sum, B. Damodaran, S. Mhaisalkar, and N. Mathews, Chem. Commun. \textbf{52}, 7118 (2016).

\bibitem{li2017cs} Z.-J. Li, E. Hofman, J. Li, A. H. Davis, C.-H. Tung, L.-Z. Wu, and W. Zheng, Adv. Funct. Mater. \textbf{28}, 1704288 (2018).

\bibitem{qiao2017} B. Qiao, P. Song, J. Cao, S. Zhao, Z. Shen, D. Gao, Z. Liang, Z. Xu, D. Song, and X. Xu, Nanotechnology \textbf{28}, 445602 (2017).

\bibitem{huang2016} H. Huang, B. Chen, Z. Wang, T. F. Hung, A. S. Susha, H. Zhong, and A. L. Rogach, Chem. Sci. \textbf{7}, 5699 (2016).

\bibitem{luo2016} B. Luo, Y.-C. Pu, S. A. Lindley, Y. Yang, L. Lu, Y. Li, X. Li, and J. Z. Zhang, Angew. Chem. \textbf{128}, 9010 (2016).

\bibitem{wang2016encap} Y. Wang, J. He, H. Chen, J. Chen, R. Zhu, P. Ma, A. Towers, Y. Lin, A. J. Gesquiere, S.-T. Wu, Y. Dong, Adv. Mater. \textbf{28}, 10710 (2016).

\bibitem{raja2016} S. N. Raja, Y. Bekenstein, M. A. Koc, S. Fischer, D. Zhang, L. Lin, R. O. Ritchie, P. Yang, and A. P. Alivisatos, ACS Appl. Mater. Interfaces \textbf{8}, 35523 (2016).

\bibitem{wei2017} Y. Wei, X. Deng, Z. Xie, X. Cai, S. Liang, P. Ma, Z. Hou, Z. Cheng, and J. Lin, Adv. Funct. Mater. \textbf{27}, 1703535 (2017).

\bibitem{sun2016ligand} S. Sun, D. Yuan, Y. Xu, A. Wang, and Z. Deng, ACS Nano \textbf{10}, 3648 (2016).

\bibitem{zhang2016enhancing} X. Zhang, H. Lin, H. Huang, C. Reckmeier, Y. Zhang, W. C. Choy, and A. L. Rogach, Nano Lett. \textbf{16}, 1415 (2016).

\bibitem{swarnkar2015colloidal} A. Swarnkar, R. Chulliyil, V. K. Ravi, M. Irfanullah, A. Chowdhury, and A. Nag, Angew. Chem. \textbf{127}, 15644 (2015).

\bibitem{xue2017constructing} J. Xue, Y. Gu, Q. Shan, Y. Zou, J. Song, L. Xu, Y. Dong, J. Li, and H. Zeng, Angew. Chem. Int. Ed. \textbf{56}, 5232 (2017).

\bibitem{chen2017solvothermal} M. Chen, Y. Zou, L. Wu, Q. Pan, D. Yang, H. Hu, Y. Tan, Q. Zhong, Y. Xu, H. Liu, B. Sun, Q. Zhang, Adv. Funct. Mater. \textbf{27}, 1701121  (2017).

\bibitem{yuandegradation} G. Yuan, C. Ritchie, M. Ritter, S. Murphy, D. E. Gomez, and P. Mulvaney, J. Phys. Chem. C. (2018).

\bibitem{huang2017lead} H. Huang, M. I. Bodnarchuk, S. V. Kershaw, M. V. Kovalenko, and A. L. Rogach, ACS Energy Lett. \textbf{2}, 2071 (2017).

\bibitem{leijtens2017towards} T. Leijtens, K. Bush, R. Cheacharoen, R. Beal, A. Bowring, and M. D. McGehee, J. Mater. Chem. A \textbf{2}, 11483 (2017).

\bibitem{haruyama2015first} J. Haruyama, K. Sodeyama, L. Han, and Y. Tateyama, J. Am. Chem. Soc. \textbf{137}, 10048 (2015).

\bibitem{kawai2015mechanism} H. Kawai, G. Giorgi, A. Marini, and K. Yamashita, Nano Lett. \textbf{15}, 3103 (2015).

\bibitem{geng2014first} W. Geng, L. Zhang, Y.-N. Zhang, W.-M. Lau, and L.-M. Liu, J. Phys. Chem. C. \textbf{118}, 19565 (2014).

\bibitem{yin2015halide} W.-J. Yin, J.-H. Yang, J. Kang, Y. Yan, and S.-H. Wei, J. Mater. Chem. A \textbf{3}, 8926 (2015).

\bibitem{iyikanat2017thinning} F. Iyikanat, E. Sari, H. Sahin, Phys. Rev. B \textbf{96}, 155442 (2017).

\bibitem{kang2017high} J. Kang and L.-W. Wang, J. Phys. Chem. Lett. \textbf{8}, 489 (2017).

\bibitem{yin2014unusual} W.-J. Yin, T. Shi, and Y. Yan, Appl. Phys. Lett. \textbf{104}, 063903 (2014).

\bibitem{kim2014role} J. Kim, S.-H. Lee, J. H. Lee, and K.-H. Hong, J. Phys. Chem. Lett. \textbf{5}, 1312 (2014).

\bibitem{azpiroz2015defect} J. M. Azpiroz, E. Mosconi, J. Bisquert, and F. De Angelis, Energy Environ. Sci. \textbf{8}, 2118 (2015).

\bibitem{haruyama2017first} J. Haruyama, K. Sodeyama, I. Hamada, L. Han, and Y. Tateyama, J. Phys. Chem. Lett. \textbf{8}, 5840 (2017).

\bibitem{bekenstein2015highly} Y. Bekenstein, B. A. Koscher, S. W. Eaton, P. Yang, and A. P. Alivisatos, J. Am. Chem. Soc \textbf{137}, 16008 (2015).

\bibitem{amgar2017} D. Amgar, A. Stern, D. Rotem, D. Porath, and L. Etgar, Nano Lett. \textbf{17}, 1007 (2017).

\bibitem{kresse1999ultrasoft} G. Kresse and D. Joubert, Phys. Rev. B \textbf{59}, 1758 (1999).

\bibitem{blochl1994projector} P. E. Bl\"{o}ch, Phys. Rev. B \textbf{50}, 17953 (1994).
 
\bibitem{kresse1993ab} G. Kresse and J. Hafner, Phys. Rev. B \textbf{47}, 558 (1993).
 
\bibitem{kresse1996efficient} G. Kresse and J. Furthm\"{u}ller, Phys. Rev. B \textbf{54}, 11169 (1996).

\bibitem{perdew1981self} J. P. Perdew and A. Zunger, Phys. Rev. B \textbf{23}, 5048 (1981).

\bibitem{ceperley1980ground} D. M. Ceperley and B. Alder, Phys. Rev. Lett. \textbf{45}, 566 (1980).

\bibitem{henkelman2006fast} G. Henkelman, A. Arnaldsson, and H. J\'{o}nsson, Computational Materials Science \textbf{36}, 354 (2006).

\bibitem{smith2015interplay} E. H. Smith, N. A. Benedek, and C. J. Fennie, Inorg. Chem. 54, 8536 (2015).

\bibitem{wang2015pressure} Y. Wang, X. Lu, W. Yang, T. Wen, L. Yang, X. Ren, L. Wang, Z. Lin, and Y. Zhao, J. Am. Chem. Soc. \textbf{137}, 11144 (2015).

\bibitem{cottingham2016} P. Cottingham and R. L. Brutchey, Chem. Commun. \textbf{52}, 5246 (2016).

\bibitem{zhang2015solution} D. Zhang, S. W. Eaton, Y. Yu, L. Dou, and P. Yang, J. Am. Chem. Soc. \textbf{137}, 9230 (2015).

\bibitem{calistru1997identification} D. M. Calistru, L. Mihut, S. Lefrant, and I. Baltog, J. Appl. Phys. \textbf{82}, 5391 (1997).

\bibitem{zhang2017zero} Y. Zhang, M. I. Saidaminov, I. Dursun, H. Yang, B. Murali, E. Alarousu, E. Yengel, B. A. Alshankiti, O. M. Bakr, and O. F. Mohammed, J. Phys. Chem. Lett. \textbf{8}, 961 (2017).

\bibitem{thumu2017mechanistic} U. Thumu, L. Houben, H. Cohen, M. Menahem, I. Pinkas, L. Avram, T. Wolf, A. Teitelboim, M. Leskes, O. Yaffe, D. Oron, M. Kazes, Chem. Mater. \textbf{30}, 84 (2017).

\bibitem{isupova1968raman} L. Isupova and E. Sobolev, J. Struct. Chem. \textbf{9}, 263 (1968).

\bibitem{willemsen1971raman} B. Willemsen, J. Inorg. Nucl. Chem. \textbf{33}, 3963 (1971).

 
 
 
 
 
 
 
 
 
\end{thebibliography}
\end{document}